\documentclass[showpacs, twocolumn]{revtex4-1}
\usepackage{graphicx}
\usepackage{amsmath}
\usepackage{appendix}
\begin{document}

\title{Dipolar Bose gas with three-body interactions at finite temperature}

\author{Abdel\^{a}ali Boudjem\^{a}a}

\affiliation{Department of Physics,  Faculty of Exact Sciences and Informatics, Hassiba Benbouali University of Chlef P.O. Box 151, 02000, Ouled Fares, Chlef, Algeria.}

%\affiliation{$^1$Department of Physics, Faculty of Sciences, Hassiba Benbouali University of Chlef P.O. Box 151, 02000, Chlef, Algeria
%\\$^2$Laboratoire de Physique Th\'{e}orique et Mod\`{e}les Statistiques, CNRS and Universit\'{e} Paris Sud, UMR8626, 91405 Orsay, France
%\\ $^3$Van der Waals-Zeeman Institute, University of Amsterdam, Science Park 904, 1098 XH Amsterdam, The Netherlands}

%\email {a.boudjemaa@univ-chlef.dz}

%\date{\today}

\begin{abstract}
We investigate effects of  three-body contact interactions on a trapped dipolar Bose gas at finite temperature
using the Hartree-Fock-Bogoliubov approximation.
We analyze numerically the behavior of the transition temperature and the condensed fraction.
Effects of the three-body interactions, anomalous pair correlations and temperature on the collective modes are discussed.

\end{abstract}


\maketitle

\section{Introduction}

Recently, Bose-Einstein condensates (BEC) with dipole-dipole interactions (DDI) have received considerable attention both experimentally and theoretically
 \cite{Baranov, Pfau,Carr, Pupillo2012}. 
Dipolar quantum gases, in stark contrast to dilute gases of isotropic interparticle interactions, offer fascinating prospects of exploring ultracold gases 
and novel many-body quantum phases with atomic interactions that are long-range and spatially anisotropic. 
Impressionnant efforts has been directed towards the ground states  properties and elementary excitations of such dipolar systems at both zero and finite temperatures
\cite {Santos, Santos1, Dell, Eberlein, Bon1, Bon, Corm, He, Biss, lime, Boudj, Boudj1, Boudj2}.

Three-body interactions (TBI) are ubiquitous and play an important role in a wide variety of interesting physical phenomena, 
and yield a new physics and many surprises not encountered in systems dominated by the two-body interactions.
Recently, many experimental and theoretical techniques have been proposed to observe  and realize the TBI in ultracold Bose gas \cite{Ham, Will, Daly, Petrov}. 
For instance, inelastic three-body processes, including observations of Efimov quantum states and atom loss from recombination have been also reported in Refs \cite{Eff, Eff1, Bed, Kra, Brut}.
Few-body forces induce also nonconventional many-body effects such as 
quantum Hall problems \cite {Grei}  and the transition from the weak to strong-pairing Abelian phase \cite {Moor, Read}.
In 2002, Bulgac \cite{Bulg} predicted  that both weakly interacting Bose and Fermi gases with attractive two-body and large repulsive TBI may form droplets.
Dasgupta  \cite{Dasg} showed that if the two-body interactions were attractive, the presence of the TBI leads to 
a nonreversible BCS-BEC crossover. 
Furthermore, many proposals dealing with effects of effective TBI in ultracold bosonic atoms in an optical lattice or a superlattice are reported in 
\cite{ Daly1,Mazz, Singh, Mahm}.
Moreover, the TBI in dilute Bose gases may give rise to considerably modify the collective excitations at both zero and finite temperatures \cite {Abdul,Hamid, Chen}, 
the transition temperature, the condensate depletion and the stability of a BEC in a one (1D)- and  two (2D)-dimensional  trapping geometry \cite {Peng, Mash}.

However, little attention has been paid to effects of TBI on dipolar BECs. 
For instance, it has been argued that TBI play a crucial role in the stablization of the supersolid state in 2D dipolar bosons \cite{Petrov1} 
and in the quantum droplet state in 3D BEC with strong  DDI \cite{Pfau1, Kui, Blakie, BoudjDp}.

Our main aim here is to study effects of the TBI on weakly interacting dipolar Bose gases in a pencake trap at finite temperature.
To this end, we employ the full Hartree-Fock-Bogoliubov (HFB) approximation.
This approach which takes into account the pair anomalous correlations has been extensively utilized 
to describe the properties of both homogenous and trapped BECs with contact interactions \cite{Hut}.
%It is worth noticing that the appearance of the anomalous correlations in the HFB theory would lead to an unphysical gap in the excitation spectrum,
%violating the Hugenholtz-Pines theorem \cite{HP}.
%The HFBP theory was shown to be a powerful tool to study the behavior of Bose gases with contact interactions at finite temperature (see for review \cite {Franc, Hut}).
We will show in particular how the interplay between the DDI, TBI and temperature can enhance the density profiles, the condensed fraction and
the collective modes of the system.

The rest of the paper is organized as follows. In Sec.\ref {3BT}, we introduce the full HFB formalism for dipolar BECs with TBI.
We discuss also the issues encountered in our model and present the resolution of these problems.
Section \ref{NR} is devoted  to presenting and discussing our numerical results.
Our conclusions are drawn in Sec.\ref{Conc}.

\section{Three-body model for dipolar bosons} \label{3BT}

Consider a dipolar BEC with contact repulsive  two-body interactions and TBI confined in a pancake-shaped trap with the dipole moments of
the particles oriented perpendicular to the plane. 
It is straightforward to check that the condensate wavefunction $\Phi ({\bf r})=\langle \hat\psi ({\bf r})\rangle$, with $\hat\psi ({\bf r})$ being 
the Bose field operator, satisfies the generalized Gross-Pitaevskii (GP) equation \cite{BoudjDp}

\begin{widetext}
\begin{align}\label{EMP}
i\hbar \frac{\partial \Phi ({\bf r},t)} {\partial t} &=\left \{ h^{sp}+ g_2 \bigg[n_c({\bf r},t) +2 \tilde n({\bf r},t) \bigg]+ \frac{g_3}{2}  \bigg [n_c^2({\bf r},t)+
 6n_c ({\bf r},t)\tilde n({\bf r},t)  + \tilde m^*({\bf r},t) \Phi^2({\bf r},t) \bigg ]  \right \} \Phi ({\bf r},t) \\
&+\left [ g_2 \tilde m ({\bf r},t) +\frac{3 g_3}{2} \tilde m({\bf r},t)  n_c ({\bf r},t) \right] \Phi^*({\bf r},t)  \nonumber \\
&+\int d{\bf r'} V_d({\bf r}-{\bf r'}) \bigg [ n ({\bf r'},t) \Phi({\bf r},t)+ \tilde n ({\bf r},{\bf r'},t)\Phi({\bf r'},t) +\tilde m ({\bf r},{\bf r'},t)\phi^*({\bf r'},t) \bigg ], \nonumber 
\end{align}
\end{widetext}
where $h^{sp} =-\hbar^2 \Delta/2m +U({\bf r})$ is the single particle Hamiltonian, $m$ is the particle mass, 
$U({\bf r})= m \omega_{\rho}^2 (\rho^2+\lambda^2 z^2)/2$, $\rho^2=x^2+y^2$,
$\lambda=\omega_{z}/\omega_{\rho}$ is the ratio between the trapping frequencies in the axial and radial directions.
The two-body coupling constant is defined by $g_2=4\pi \hbar^2 a/m$ with $a$ being  the $s$-wave scattering length which can be adjusted using a magnetic Feshbach resonance.
The three-body coupling constant $g_3$ is in general a complex number with $Im(g_3)$ describing the three-body recombination loss
and $Re(g_3)$  accounting for the three-body scattering parameter.
In the present paper, we do not consider the three-body recombination terms i.e. $Im(g_3)=0$, so the system is stable which is consistent with recent experiments
\cite{Kra, Pfau1}.
The DDI potential is $V_d({\bf r}) = C_{dd} (1-3\cos^2\theta) / (4\pi r^3)$,
where $C_{dd} ={\cal M}_0 {\cal M}^2 (= d^2/\epsilon_0)$ is the magnetic (electric) dipolar interaction strength, 
and $\theta$ is the angle between the relative position of the particles ${\bf r}$ and the direction of the dipole.
The condensed and noncondensed densities are defined, respectively as
$n_c({\bf r})=|\Phi({\bf r})|^2$, $\tilde n ({\bf r})= \langle \hat {\bar\psi}^\dagger ({\bf r}) \hat {\bar\psi} ({\bf r}) \rangle $ and 
 $n({\bf r})=n_c({\bf r})+\tilde n ({\bf r})$ is the total density.
The term $\tilde n ({\bf r, r'})$ and $\tilde m ({\bf r, r'})$ are respectively, the normal and the anomalous one-body density matrices
which account for the dipole exchange interaction between the condensate and noncondensate.

Equation (\ref{EMP}) describes the coupled dynamics of the condensed and noncondensed components.
For $g_3=0$, it recovers the generalized nonlocal finite-temperature GP equation with two-body interactions.
For $\tilde m =0$, Eq.(\ref{EMP}) reduces to the HFB-Popov equation \cite{Bon, Corm,BoudjDp} which is gapless theory.
For $\tilde m=\tilde n =0$, it reduces to standard GP equation that describes dipolar Bose gases only at zero temperature.

Upon linearizing Eq.(\ref{EMP})  around a static solution $\Phi_0$, utilizing the parameterization
$\Phi({\bf r},t)=[\Phi_0({\bf r})+\delta \Phi({\bf r},t) ] e^{-i\mu t/\hbar}$, 
where $\delta \Phi = \sum_k [u_k ({\bf r}) e^{-i \varepsilon_k t/\hbar}+ v_k({\bf r}) e^{i \varepsilon_k t/\hbar}] $, 
and $\varepsilon_k$ is the Bogoliubov excitations energy.
The quasi-particle amplitudes $ u_k({\bf r}), v_k({\bf r}) $ satisfy the generalized nonlocal Bogoliubov-de-Gennes (BdG) equations \cite{BoudjDp}:
\begin{widetext}
\begin{align}
\varepsilon_k u_k ({\bf r}) &= \hat {\cal L} u_k ({\bf r})+ \hat {\cal M} v_k ({\bf r}) + \int d{\bf r'} V_d({\bf r}-{\bf r'}) n ({\bf r},{\bf r'}) u_k ({\bf r'}) 
+ \int d {\bf r'}  V_d({\bf r}-{\bf r'}) \bar m  ({\bf r},{\bf r'}) v_k ({\bf r'}), \label{BdG1} \\ 
-\varepsilon_k v_k ({\bf r}) &= \hat {\cal L} v_k ({\bf r})+ \hat {\cal M} u_k ({\bf r}) + \int d{\bf r'} V_d({\bf r}-{\bf r'}) n ({\bf r},{\bf r'}) v_k ({\bf r'}) 
+ \int d {\bf r'}  V_d({\bf r}-{\bf r'}) \bar m  ({\bf r},{\bf r'})  u_k ({\bf r'}), \label{BdG2}
\end{align}
\end{widetext}
where 
$\hat {\cal L}=h^{sp}+ 2g_2n ({\bf r})+ 3g_3 [n_c^2 ({\bf r}) +4 n_c ({\bf r}) \tilde n ({\bf r})+\tilde m^* ({\bf r}) \Phi^2({\bf r})+\tilde m ({\bf r}) \Phi^{*2}({\bf r})]/2 
+ \int d{\bf r'} V_d({\bf r}-{\bf r'}) n ({\bf r'})-\mu$, 
$\hat {\cal M}=g_2 [\Phi_0^2({\bf r})+ \tilde m ({\bf r})]+g_3[ n_c^2({\bf r})+3\Phi_0^2({\bf r}) \tilde n ({\bf r})+3\Phi_0^2({\bf r}) \tilde m ({\bf r}) ]$,
$n ({\bf r},{\bf r'})= \Phi_0^*({\bf r'}) \Phi_0({\bf r})+ \tilde n ({\bf r},{\bf r'})$ and $\bar m  ({\bf r},{\bf r'})= \Phi_0({\bf r'}) \Phi_0({\bf r}) +\tilde m  ({\bf r},{\bf r'})$. \\
Equations (\ref{BdG1}) and (\ref{BdG2}) describe the collective excitations of the system.
The normal and the anomalous one-body density matrices  can be obtained  employing the transformation
 $\hat {\bar\psi}=\sum_k [u_k ({\bf r}) \hat b_k+ v_k^*({\bf r}) \hat b_k^\dagger] $ 
\begin{align} 
 \tilde n ({\bf r, r'})&= \sum_k \bigg\{ \left[u_k^*({\bf r'}) u_k ({\bf r})+v_k({\bf r'})v_k^*({\bf r}) \right] N_k({\bf r})  \label {HFB1}  \\
&+v_k({\bf r'})v_k^*({\bf r})\bigg\},  \nonumber \\
 \tilde m ({\bf r, r'})&= -\sum_k \bigg\{ \left[u_k({\bf r'}) v^*_k ({\bf r})+u_k({\bf r})  v_k^*({\bf r'}) \right] N_k({\bf r})  \label {HFB2}  \\
&+u_k({\bf r'})v_k^*({\bf r})\bigg\},  \nonumber 
\end{align}
where $N_k=\langle \hat b_k^{\dagger} \hat b_k\rangle=[\exp(\varepsilon_k/T)-1]^{-1}$ are occupation numbers for the excitations.  
The noncondensed and anomalous densities can simply be obtained  
by setting, respectively $ \tilde n ({\bf r})=\tilde n ({\bf r, r})$ and $ \tilde m ({\bf r})=\tilde m ({\bf r, r})$ in Eqs.(\ref{HFB1}) and (\ref{HFB2}).

From now on we assume that $\tilde n ({\bf r},{\bf r'})=\tilde m ({\bf r},{\bf r'})=0$ for ${\bf r} \neq {\bf r'}$ \cite{Bon,Bon1}.
It is worth stressing that the omission of the long-range exchange term does not preclude the stability of the system \cite {Bon, He, Biss, Zhan, Bail, Tick, BoudjDp}.

As is well known, the full HFB theory sustains some hindrances notably the appearence of 
an unphysical gap in the excitation spectrum and the divergence of the anomlaous density.
In fact, this violation of the conservation laws in the HFB theory is due to the inclusion of the anomalous density which in general leads
to a double counting of the interaction effects.
The common way to circumvent this problem is to neglect $\tilde m$ in the above equations, which restores the symmetry and hence leads to a gapless theory,
but this is nothing else than the Popov approximation. To go consistently beyond the Popov theory, one should renormalize the coupling constant 
taking into account many-body corrections for scattering between the condensed atoms on one hand and the condensed and thermal atoms on the other.
Following the procedure outlined in Refs \cite{Davis, Morgan, Boudj8, Boudj9, Boudjbook} we obtain
\begin{align} \label{Renor}
&g_2 |\Phi|^2\Phi+g_2\tilde{m}\Phi^*+\frac{3g_3 }{2} n_c\tilde{m}\Phi^*\\
& =g_2 \bigg[1+\frac{\tilde {m}(1+3g_3n_c/g_2) }{\Phi ^2}\bigg] |\Phi|^2\Phi \nonumber\\
&= g_R |\Phi|^2\Phi \nonumber.
\end{align} 
This spatially dependent effective interaction, $g_R$ is somehow equivalent to the many body $T$-matrix \cite{Hut}.  
A detailed derivation of $g_R$, including the term $g_3$, will be given elsewhere. 
It is easy to check that if one substitutes (\ref{Renor}) in the HFB equations, we therefore, reinstate the gaplessness of the spectrum and the convergence of the anomalous density.
%A similar technique has been employed to calculate  many-body corrections to the mean-field interaction potential using the classical-field formalism \cite{Wright}. 

\section{Numerical results} \label{NR}

%We have solved Eqs.(\ref{EMP})-(\ref{Renor}) iteratively and in a self-consistent way assuming that $\tilde n ({\bf r},{\bf r'})=0$ for ${\bf r} \neq {\bf r'}$ \cite{Bon,Bon1}.
%It has been shown that the omission of the long range exchange term do not qualitatively affact the stability of the system \cite {Bon, He, Biss, Zhan, Bail, Tick, BoudjDp}.

For numerical purposes, it is useful to set the  Eqs.(\ref{EMP})-(\ref{Renor}) into a dimensionless from. We introduce the following dimensionless parameters:
the relative strength $\epsilon_{dd}=C_{dd}/3g_2 $ ($\epsilon_{dd}=0.16$ for Cr atoms) 
which describes the interplay between the DDI and short-range interactions, and $\bar g_3 =g_3 n_c/g_2$ describes the ratio between the two-body interactions and TBI.
Throughout the paper, we express lengths and energies in terms of the transverse harmonic oscillator length $l_0=\sqrt{\hbar/m \omega_{\rho}}$
and the trap energy $\hbar \omega_{\rho}$, respectively. 

\begin{figure}
\includegraphics[scale=0.45]{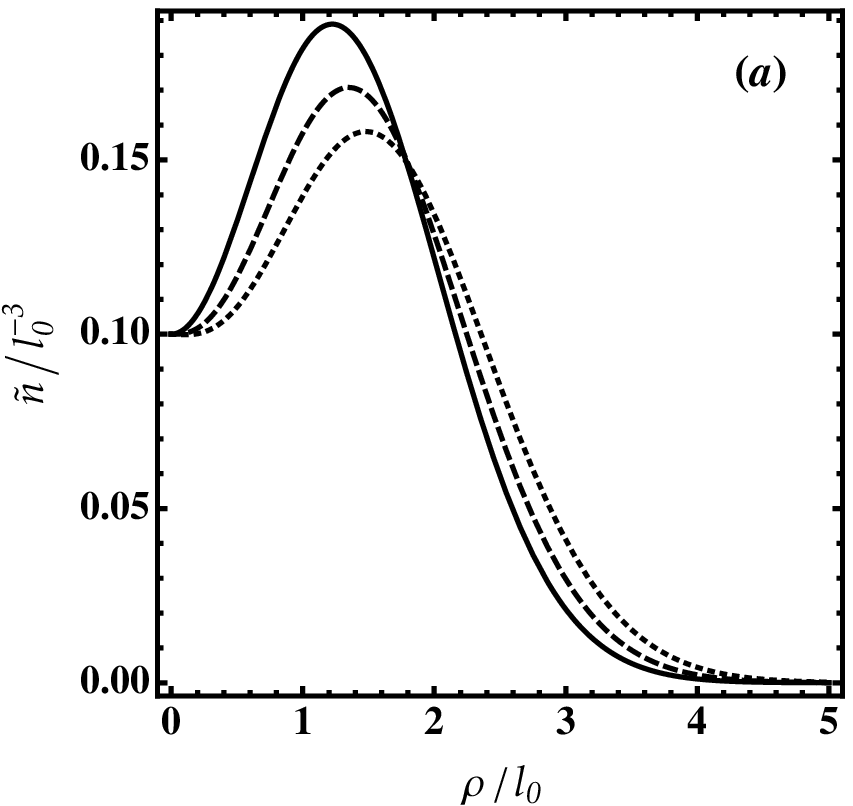}
\includegraphics[scale=0.45]{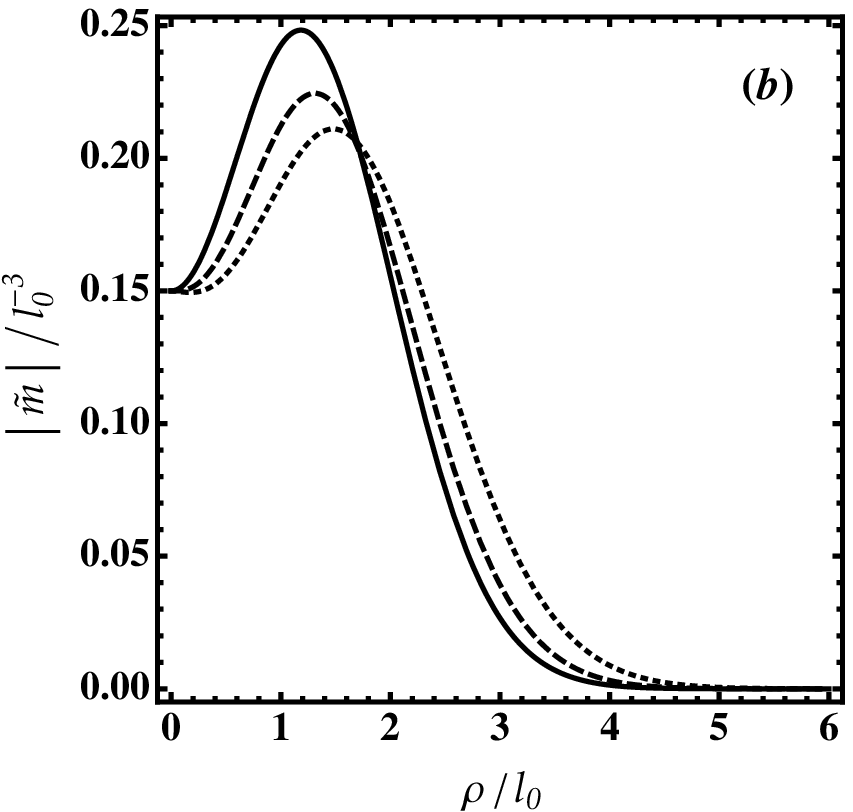}
\caption { (Color online)  Density profiles of the noncondensed (a) and anomalous (b) components for several values of the three-body interactions $\bar g_3$ at $T/T_c=0.2$.
Solid line: $\bar g_3=0.05$, dashed line: $\bar g_3=0.055$ and dotted line: $\bar g_3=0.06$.
Parameters are : $N=10^5$ of ${}^{52}$Cr atoms, $\lambda=4$, and the scattering length $a=20 a_0$.
Here the contact interactions can be tuned using the Feshbach resonance making the dipolar strength more lager.}
\label{dens}
\end{figure}

Figure.\ref{dens} shows that the noncondensed and the anomalous densities increase with the TBI which leads to reduce the condensed density.
A careful observation of the same figure reveals that the $\tilde m$ is larger than $\tilde n$ at low temperature which is in fact natural since the anomlaous density 
itself arises and grows with interactions  \cite{Boudj2011, Boudj2012}. When the temperature approaches to the transition, one can expect that $\tilde m$ 
vanishes similar to the case of a BEC with a pure contact interaction \cite{Boudj2011, Boudj2012}.

\begin{figure}[htb] 
\includegraphics[scale=0.8]{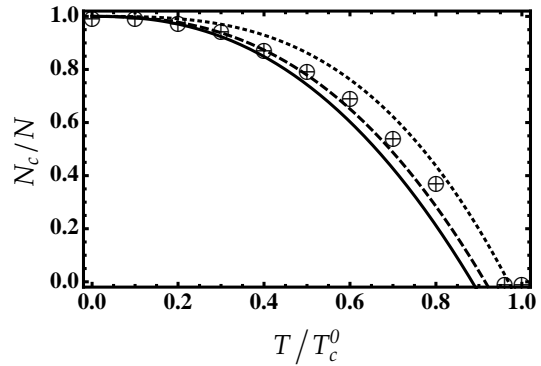}
\caption {Condensate fraction as a function of reduced temperature $T/T_c^0$ (where $T_c^0$ is the ideal gas critical temperature). 
Solid line: full HFB with $\bar g_3=0.05$,  dashed line: full HFB with $\bar g_3=0$, circles: HFB-Popov approximation without TBI, and dotted line: ideal gas $N_c/N=1-(T/T_c^0)^3$. 
Parameters are the same as in Fig.\ref{dens}.}
\label{Frac}
\end{figure}
In Fig.\ref{Frac} we compare our prediction for the condensed fraction $N_c/N$ with the HFB-Popov theoretical treatment and the noninteracting gas.
As is clearly seen, our results diverge from those of the previous approximations due to the effects of the TBI.
This means that both the condensed fraction and the transition temperature decrease with increasing the TBI.

\begin{figure}[ htb] 
\includegraphics[scale=0.8]{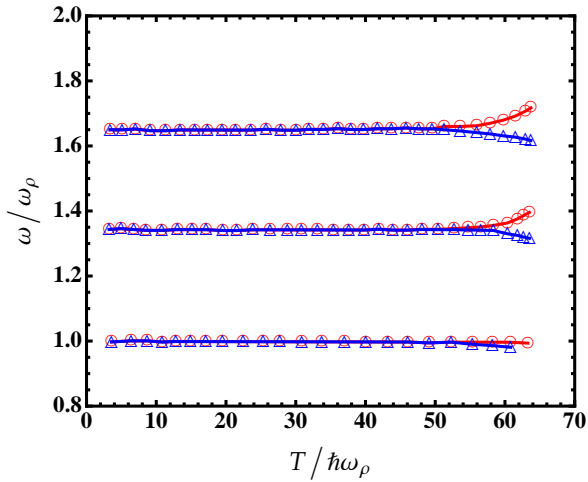}
\caption { (Color online) Lowest mode frequencies of ${}^{52}$Cr  BEC  as a function of temperature.
Triangles: our predictions, circles: HFB-Popov approximation without TBI. 
For each angular momentum number $m=0,1,2$, we plot only the lowest mode.
Parameters are the same as in Fig.\ref{dens}.}
\label{Exc}
\end{figure}

Before leaving this section, let us unveil the role of the TBI on the collective excitations.
According to Fig.\ref{Exc}, we can observe that our results deviate from the HFB-Popov (which have also been found in \cite{Bon}) 
at higher temperatures for $m=0$ and $2$ excitations. 
The reason of such a downward shift which enlarges as the temperature approaches $T_c$, is the inclusion of both anomalous pair correlations and TBI.
A similar behavior holds in the case of BEC with short-range interactions (see e.g. \cite{Hut}).
Fig.\ref{Exc} depicts also that both the full HFB  and the HFB-Popov produce a small shift from 1 for the Kohn mode $\omega/\omega_{\rho}=1$ at higher temperatures.
One possibility to fix this problem might be the inclusion of the dynamics of the noncondensed and the anomalous components.
A suitable formalism to explore such a dynamics is the time-dependent HFB theory \cite{Boudj8, Boudj9, Boudjbook, Boudj2011, Boudj2012}.

\section{Conclusion} \label{Conc}

In conclusion, we have deeply investigated the properties of dipolar Bose gas confined by a cylindrically symmetric harmonic trapping potential
in the presence of TBI at finite temperature.
The numerical simulation of the full HFB model emphasized that the condensed fraction and the transition temperature 
are reduced by the TBI.
Effects of the TBI and temperature on the collective modes of the system are notably highlighted.
We found that the full HFB approach in the presence of the TBI reproduces the HFB-Popov results of \cite{Bon} only at low temperature
while both approaches diverge each other when the temperature is close to $T_c$.
One can expect that the same behavior persists in the case of a density-oscillating ground states known as a biconcave state predicted in Ref \cite{Bon1}.

%We have shown also that under certain condition of interaction and trap geometry, both the condensate and the thermal cloud develop  a deep local minimum
%at the trap center owing to the repulsive three-body forces. 
%The temperature dependence of the contrast of such biconcave state is notably discussed.

\section*{Acknowledgments}
We are grateful to Dmitry Petrov and Axel Pelster for the careful reading of the manuscript and helpful comments.


\begin{thebibliography}{28}

\bibitem{Baranov} See for review:  M. A. Baranov, Physics Reports {\bf 464}, 71 (2008).
\bibitem{Pfau}  See for review: T. Lahaye et al., Rep. Prog. Phys. {\bf 72}, 126401 (2009).
\bibitem{Carr} See for review: L.D. Carr, D. DeMille, R.V. Krems, and J. Ye, New J. Phys.  {\bf 11},  055049 (2009).
\bibitem{Pupillo2012} See for review: M.A. Baranov, M. Delmonte, G. Pupillo, and P. Zoller, Chemical Reviews, {\bf 112}, 5012 (2012).
\bibitem{Santos} L. Santos, G. V. Shlyapnikov, P. Zoller, M. Lewenstein, Phys. Rev. Lett. {\bf 85}, 3745 (2000).
\bibitem {Santos1} L. Santos, G. V. Shlyapnikov, and M. Lewenstein, Phys. Rev. Lett. {\bf 90}, 250403 (2003).
\bibitem {Dell} D. H. J. O’Dell, S. Giovanazzi, and C. Eberlein, Phys. Rev. Lett.{\bf 92}, 250401 (2004).
\bibitem{Eberlein} C.Eberlein, S.Giovanazzi, D.H J O'Dell, Phys. Rev. A {\bf 71}, 033618 (2005). 
\bibitem {Bon1}  S. Ronen, D. C. E. Bortolotti, and J. L. Bohn, Phys. Rev. A {\bf 74}, 013623 (2006).
\bibitem {Bon}  S. Ronen, D. C. E. Bortolotti, and J. L. Bohn, Phys. Rev. A. {\bf 76}, 043607 (2007).
\bibitem {Corm}  S. C. Cormack and D. A. W. Hutchinson, Phys. Rev. A {\bf 86}, 053619 (2012).
\bibitem {He} L. He, J.-N. Zhang, Y. Zhang, and S. Yi, Phys. Rev. A {\bf 77}, 031605 (2008).
\bibitem {Biss} R. N. Bisset, D. Baillie, and P. B. Blakie, Phys. Rev. A {\bf 86}, 033609 (2012). 
\bibitem {lime} Aristeu R. P. Lima and Axel Pelster, Phys. Rev. A {\bf 84}, 041604 (R) (2011); Phys. Rev. A {\bf 86}, 063609 (2012).
\bibitem{Boudj}  A. Boudjemaa and G.V. Shlyapnikov, Phys. Rev. A {\bf 87}, 025601 (2013).
\bibitem{Boudj1} A. Boudjem\^{a}a, J. Phys. B: At. Mol. Opt. Phys.  {\bf 48}, 035302 (2015).
\bibitem{Boudj2} A. Boudjem\^{a}a, J. Phys. A: Math. Theor. {\bf 49}, 285005 (2016).

\bibitem{Ham}  See for review: H.-W. Hammer, A. Nogga, and A. Schwenk, Rev. Mod. Phys. {\bf 85}, 197 (2013).
\bibitem{Will}  S. Will, T. Best, U. Schneider, L. Hackerm\"uller, D. S. L\"uhmann, I. Bloch, Nature (London) {\bf 465}, 197 (2010).
\bibitem{Daly}  A. J. Daley, J. Simon, Physical Review A {\bf 89},  053619 (2014).
\bibitem{Petrov}  D. S. Petrov, Phys. Rev. Lett. {\bf 112}, 103201  (2014).
\bibitem{Eff}  V. Efimov Phys. Lett. B {\bf 33}, 563 (1970).
\bibitem{Eff1} V. Efimov,  Sov. J. Nucl. Phys. {\bf 12} 589 (1971).
\bibitem{Bed}  P F Bedaque, E. Braaten and H-W-Hammer, Phys. Rev. Lett. {\bf 85} 908 (2000 ).
\bibitem{Kra}  T. Kraemer et {\textit al}. Nature {\bf 440}, 315 (2006). 
\bibitem{Brut}  E. A. Burt,  R. W. Ghrist, C. J. Myatt, M J. Holland, E. A. Cornell and C.E. Wieman, Phys. Rev. Lett. {\bf 79} 337 (1997).
\bibitem{Grei} Martin Greiter, Xiao-Gang Wen, and Frank Wilczek, Phys. Rev. Lett. {\bf 66}, 3205 (1991).
\bibitem{Moor}  G. Moore and N. Read, Nucl. Phys. B {\bf 360}, 362 (1991).
\bibitem{Read}  N. Read and D. Green, Phys. Rev. B {\bf 61}, 10267 (2000).
\bibitem{Bulg} A. Bulgac, Phys. Rev. Lett. {\bf 89}, 050402 (2002).
\bibitem{Dasg}  R. Dasgupta, Phys. Rev. A {\bf 82}, 063607 (2010).
\bibitem{Mash}  H. P. B\"uchler, A. Micheli, and P. Zoller, Nat. Phys. {\bf 3}, 726 (2007).
\bibitem{Daly1} A. J. Daley, J.M. Taylor, S. Diehl, M. Baranov, and P. Zoller, Phys. Rev. Lett. {\bf 102}, 040402 (2009).
\bibitem{Mazz}  L. Mazza, M. Rizzi, M. Lewenstein, and J. I. Cirac, Phys. Rev. A {\bf 82}, 043629 (2010).
\bibitem{Singh} M. Singh, A. Dhar, T. Mishra, R. V. Pai, B. P. Das, Phys. Rev. A {\bf 85}, 051604 (2012).
\bibitem{Mahm}  K.W. Mahmud and E. Tiesinga, Phys. Rev. A {\bf 88}, 023602 (2013).
\bibitem{Abdul} F.K. Abdullaev, A. Gammal, L. Tomio and T. Frederico, Phys. Rev. A {\bf 63} 043604 (2001).
\bibitem{Hamid} Hamid Al-Jibbouri, Ivana Vidanovic, Antun Balaz, and Axel Pelster, J. Phys. B: At. Mol. Opt. Phys. {\bf 46},  065303 (2013).
\bibitem{Chen}  H-C Li, K-J. Chen and J-K  Xue, Chin. Phys. Lett. {\bf 27} 030304 (2010)
\bibitem{Peng} Peng P and Li G-Q  Chin. Phys. B {\bf 18}, 3221 (2009).
\bibitem{Mash}  M. S. Mashayekhi, J.-S. Bernier, D. Borzov, J.-L. Song, and F. Zhou, Phys. Rev. Lett. {\bf 110}, 145301 (2013).
\bibitem{Petrov1}  Zhen-Kai Lu, Yun Li, D. S. Petrov, and G. V. Shlyapnikov, Phys. Rev. Lett. {\bf 115}, 075303 (2015).
\bibitem{Pfau1} H. Kadau, M. Schmitt, M. Wenzel, C. Wink, T. Maier, I. Ferrier-Barbut and T. Pfau, Nature {\bf 530} ,194 (2016).
\bibitem{Kui} Kui-Tian Xi and Hiroki Saito, Phys. Rev. A {\bf 93}, 011604(R) (2016).
\bibitem{Blakie} P. B. Blakie, Phys. Rev. A {\bf 93}, 033644 (2016).
\bibitem{BoudjDp}  A. Boudjem\^{a}a, Annals of Physics, {\bf 381}, 68 (2017).
\bibitem{Hut} See for review: D. A. W. Hutchinson, R. J. Dodd, K. Burnett, S. A. Morgan, M. Rush, E. Zaremba, N. P. Proukakis, M. Edwards, and C. W. Clark, J. Phys. B {\bf 33}, 3825 (2000).
\bibitem {Zhan}  J.-N. Zhang and S. Yi, Phys. Rev. A {\bf 81}, 033617 (2010).
\bibitem {Bail}  D. Baillie and P. B. Blakie, Phys. Rev. A {\bf 82}, 033605 (2010).
\bibitem {Tick} C. Ticknor, Phys. Rev. A {\bf 85}, 033629 (2012).

%\bibitem{HP} N. M. Hugenholtz and D. Pines, Phys. Rev. {\bf 116}, 489 (1959).
%\bibitem{Franc} Franco Dalfovo, Stefano Giorgini, Lev P. Pitaevskii, and Sandro Stringari, Rev. Mod. Phys. {\bf 71}, 463 (1999).
\bibitem{Morgan}  S. A. Morgan, J. Phys. B {\bf 33}, 3847 (2000).
\bibitem{Davis}   M. J. Davis, S. A. Morgan, and K. Burnett, Phys. Rev. Lett. {\bf 87}, 160402 (2001).
\bibitem {Boudj8} A. Boudjem\^{a}a, Phys. Rev. A {\bf 88}, 023619 (2013).
\bibitem {Boudj9} A. Boudjem\^{a}a, J. Phys. A: Math. Theor. {\bf 48}  045002 (2015).
\bibitem {Boudjbook} A. Boudjem\^{a}a, Degenerate Bose Gas at Finite Temperatures, LAP LAMBERT Academic Publishing (2017).
\bibitem{Boudj2011} A. Boudjem\^{a}a and M. Benarous, Phys. Rev. A {\bf 84}, 043633 (2011).
\bibitem{Boudj2012} Abdel\^{a}ali Boudjem\^{a}a, Phys. Rev. A {\bf 86}, 043608 (2012).





\end{thebibliography}
\end{document}